\documentclass[aps,amssymb,showpacs,supercriptaddress,floatfix,preprint, longbibliography]{revtex4-2}
\usepackage{amssymb}
\usepackage{amsmath}
\usepackage{graphicx}
\usepackage{bm}
\usepackage[normalem]{ulem}
\usepackage{caption}
\usepackage{subcaption}
\usepackage{bm}
\usepackage{color}
\usepackage[titletoc]{appendix}

\begin{document}
\graphicspath{{images/}}
\newcommand{\change}[1]{{\color{red}{#1}}}
\newcommand{\dchange}[1]{{\color{blue}{#1}}}


\preprint{AIP/123-QED}

\title{Device-area selection of memristive transport regimes in epitaxial $Hf_{0.5}Zr_{0.5}O_{2}$-based ferroelectric devices}

\author{Priscila A. Tapia Presas$^{1,\dagger}$, Lautaro Galarregui$^{1,\dagger}$, Wilson Román Acevedo$^{1,2}$, Myriam H. Aguirre$^{3,4,5}$, José Santiso$^{6}$, Sylvia Matzen$^{7}$, Beatriz Noheda$^{8,9}$, Diego Rubi$^{1}$}

\affiliation{$^{1}$Laboratorio de Ablación Láser, INN-CONICET-CNEA, Gral. Paz 1499, 1650 San Martín, Argentina \\ $^{2}$ Departamento de Ciencias Básicas, Facultad de Ingeniería y Ciencias Exactas, Universidad Argentina de la Empresa (UADE), Lima 717 (1073) Buenos Aires, Argentina \\ $^{3}$ Instituto de Nanociencia y Materiales de Aragón (INMA-CSIC), Campus Rio Ebro C/Mariano Esquillor s/n, 50018 Zaragoza, Spain \\ $^{4}$ Dpto. de Física de la Materia Condensada, Universidad de Zaragoza, Pedro Cerbuna 12, 50009 Zaragoza, Spain \\ $^{5}$ Laboratorio de Microscopías Avanzadas, Edificio I+D, Campus Rio Ebro C/Mariano Esquillor s/n, 50018 Zaragoza, Spain \\ $^{6}$ Catalan Institute of Nanoscience and Nanotechnology (ICN2), Campus UAB, Bellaterra, Barcelona 08193, Spain \\ $^{7}$ Centre de Nanosciences et de Nanotechnologies, CNRS, Université Paris-Saclay, Palaiseau, 91120 France \\ $^{8}$ CogniGron—Groningen Cognitive Systems and Materials Center, University of Groningen, Nijenborgh 4,
9747AG Groningen, The Netherlands \\ $^{9}$ Zernike Institute for Advanced Materials, University of Groningen, Nijenborgh 4, 9747AG Groningen, The Netherlands }

\thanks{$\dagger$ These authors contributed equally to this work.}

\email{Corresponding author; email: diego.rubi@gmail.com}

\begin{abstract}

Ferroelectric memristive devices based on hafnia are promising systems for neuromorphic electronics, yet the interplay between polarization-modulated resistive changes and defect-mediated transport often leads to complex and debated switching mechanisms. Here, we investigate this competition in epitaxial Hf$_{0.5}$Zr$_{0.5}$O$_2$/La$_{0.67}$Sr$_{0.33}$MnO$_3$ heterostructures with Pt top electrodes, by combining structural, ferroelectric, and memristive characterization with a statistical analysis across a broad range of device areas spanning three orders of magnitude. We identify two distinct memristive regimes with opposite resistance–voltage chiralities. Small devices exhibit a low-resistance state that scales inversely with area, consistent with area-distributed electronic transport, while larger devices display an area-independent resistance indicative of localized conductive channels. A statistical nucleation model quantitatively captures this behavior and yields a crossover characteristic area $A^* \approx 10^3$ $\mu$m$^2$. This crossover also correlates with the onset of ferroelectric wake-up for the larger devices, linking conductive-channel nucleation and oxygen-vacancy redistribution within a unified physical picture. These results establish lateral device size as a key parameter controlling the dominant transport mechanism in epitaxial hafnia-based devices.

\end{abstract}

\maketitle

The rapid growth of artificial intelligence and data-intensive computing is pushing conventional von Neumann architectures toward fundamental limits in energy efficiency and bandwidth \cite{Mehonic2022}. Neuromorphic hardware \cite{Mehonic2024Roadmap,Zolf24,Aguirre2024MemristorANN,Schmidgall2024BrainInspired}, inspired by biological neural networks, has therefore emerged as a promising route to integrate memory and computation within the same physical elements by exploiting devices with continuously tunable conductance.

Memristive devices based on transition-metal oxides are among the most widely investigated building blocks for such hardware because their resistance can be reversibly modified by electrical stress \cite{saw_2008,yu_2017}. In many oxide systems, resistive switching originates from oxygen-vacancy migration \cite{roz_2010, herpers}, enabling large resistance contrasts but also introducing stochastic filament formation \cite{degraeve2015} and device-to-device variability \cite{rieck_21}.

Ferroelectric memristive devices offer an alternative in which resistance switching is governed by polarization reversal. In ultrathin ferroelectric tunnel junctions (FTJs) \cite{tsy_2006,gar_2009,ye_2009,chan_2012}, polarization modulates the tunneling barrier, whereas in thicker structures it controls interfacial Schottky barriers which control the device resistance \cite{blom_1994, mey_2006, pin_2010, Ferreyra2020, rengifo_2022}.

The discovery of robust ferroelectricity in hafnia-based oxides \cite{boscke_2011} has greatly expanded the technological relevance of ferroelectric memdevices. Their CMOS compatibility, nanoscale ferroelectric stability, and scalability \cite{Schroeder2019,Noheda2023LessonsHfO2} have made hafnia-based systems a central platform for neuromorphic devices \cite{Brivio_2022}. At the same time, oxygen-vacancy migration \cite{nukala_2021,sul_2019,Lee2023OxygenVacanciesHfO2}, interfacial effects \cite{Liu2024}, and field-driven structural changes \cite{Jan2023OperandoHZO} can coexist with polarization switching and strongly influence transport, particularly in epitaxial Hf$_{0.5}$Zr$_{0.5}$O$_2$/La$_{0.67}$Sr$_{0.33}$MnO$_3$/SrTiO$_3$ (HZO/LSMO/STO) heterostructures with Pt top electrodes \cite{wei_2018,nukala_2021,sul_2019}.

Although switching between ferroelectric and filamentary regimes has been demonstrated in hafnia devices \cite{knabe_2023, Long2023FERBreakdown}, comparatively little attention has been paid to whether the competition between barrier-controlled transport and localized defect-dominated conduction depends on lateral device size. Establishing this dependence is essential both for clarifying the microscopic origin of resistive states and for defining device dimensions that preserve spatially homogeneous and reproducible operation.

Here, we investigate memristive switching in Pt/HZO/LSMO junctions with a $\sim$6-nm ferroelectric layer across devices with different lateral areas. We identify two regimes with opposite hysteresis chiralities and distinct scaling laws for $R_{\mathrm{low}}$: small devices predominantly follow $R_{\mathrm{low}} \propto 1/A$, whereas larger devices progressively develop area-independent resistance. A Poisson nucleation picture captures this crossover.

HZO/LSMO epitaxial thin films were grown on (001) STO single crystals by Pulsed Laser Deposition (PLD) assisted by Reflection High-Energy Electron Diffraction (RHEED). A solid-state Nd:YAG pulsed laser with a wavelength of 266 nm was used. The deposition temperature was set at 825 °C and 850 °C for LSMO and HZO, respectively, while the oxygen background pressure was set at 0.3 mbar for LSMO and in the range 0.005-0.3 mbar for HZO. The thickness of the layers was estimated from x-ray reflectivity measurements as $\approx$ 30 nm for LSMO and 5.2-8.1 nm for HZO (thickness decreases as the oxygen pressure increases, for the same number of pulses), as shown in Fig. S1. Pt top electrodes were microfabricated by combining optical lithography and sputtering. Two device geometries were implemented: circular electrodes with radii in the range 45–500~$\mu$m, and square junctions with smaller lateral areas (lateral sizes between 10–200~$\mu$m) connected to a larger contact pad for electrical probing. In the latter configuration, the contact pad is electrically isolated from the HZO layer by a SiN insulating layer. Both geometries are shown in Fig.~S2 of the Supplementary Material. 

The structural properties of the HZO/LSMO/STO heterostructures were investigated by x-ray diffraction (XRD). Fig.~1(a) shows the $\theta$--$2\theta$ XRD patterns of HZO films grown under different oxygen background pressures. In addition to the substrate reflections, distinct peaks associated with the HZO layer are observed, confirming the formation of a crystalline film for oxygen pressures $\geq$ 0.1 mbar. The position of these reflections suggests the coexistence of the (111)-oriented polar rhombohedral (r) phase and the (001)-oriented non-polar monoclinic (m) phase, which corresponds to the lowest-free-energy structure of HZO \cite{boscke_2011,Noheda2023LessonsHfO2}. Laue oscillations around the r-HZO peak are indicative of the high crystal quality of the films, with separation in agreement with film thickness. As the oxygen pressure increases, the monoclinic contribution progressively diminishes and reduces to a very weak, barely distinguishable feature from the background for growth oxygen pressures $\geq$ 0.2~mbar, so that the films can be regarded as predominantly rhombohedral within the experimental detection limits. We will focus on these samples, displaying thicknesses $\lessapprox$ 6 nm, from now on.


This phase assignment is further supported by an x-ray diffraction map recorded around the (202) HZO in-plane reflection, shown in Fig.~1(b). To improve the signal-to-background ratio in these ultrathin films, we used an in-plane grazing-incidence geometry and mapped the diffracted intensity as a function of $2\theta$ and azimuthal angle $\phi$. The resulting map displays twelve symmetrically distributed diffraction maxima, consistent with the multiplicity expected for the rhombohedral phase grown on (001) LSMO, thereby providing strong evidence for its stabilization in the HZO layer, as previously reported for HZO films \cite{wei_2018, petraru24} and related fluorite-like polar oxides such as ZrO$_2$ \cite{elBoutaybi2022ZrO2rphase}.

The epitaxial quality of the heterostructure was further assessed by reflection high-energy electron diffraction (RHEED) during growth. Representative RHEED patterns, shown in Fig.~1(c), display streaky diffraction features characteristic of a smooth and well-ordered surface (confirmed by atomic force microscopy topographies with extracted root mean square roughness $\lessapprox$ 1 nm, not shown here), confirming the epitaxial growth of both the LSMO electrode and the HZO layer.


Cross-sectional transmission electron microscopy (TEM) was used to investigate the structural quality of the heterostructures. High-angle annular dark-field scanning transmission electron microscopy (HAADF-STEM) images were acquired using a FEI Titan G2 microscope equipped with a probe corrector. The TEM cross-section displayed in Fig.~1(d), corresponding to a heterostructure grown under 0.2 mbar oxygen pressure, reveals a sharp interface between the HZO layer and the LSMO electrode together with the presence of well defined crystalline and highly oriented planes in the HZO layer.


Together, these structural characterizations confirm the epitaxial growth of the HZO/LSMO/STO heterostructures and the stabilization of a polar HZO phase, providing a suitable platform to investigate both ferroelectric and memristive properties.

To assess the ferroelectric character of the HZO layer, local piezoresponse force microscopy (PFM) measurements were performed using a Cypher system (Oxford Instruments). Figs.~2(a) and (b) show the PFM amplitude and phase images obtained after poling the HZO film by writing square domain patterns with a biased conductive tip (-6 V for the outer square and +6 V for the inner one), while the bottom LSMO electrode was grounded. The amplitude image displays no significant contrast, except in the regions corresponding to domain walls, whereas the phase image reveals a clear contrast between adjacent regions. This behavior demonstrates the presence of switchable polarization in the HZO layer. Additional dual AC resonance tracking (DART) PFM measurements, presented in Fig.~S3 of the Supplementary Material, further confirm the ferroelectric character of the HZO layer through the characteristic phase reversal and butterfly-shaped amplitude response.

\begin{figure*}[]
\centering
\includegraphics[scale=0.6]{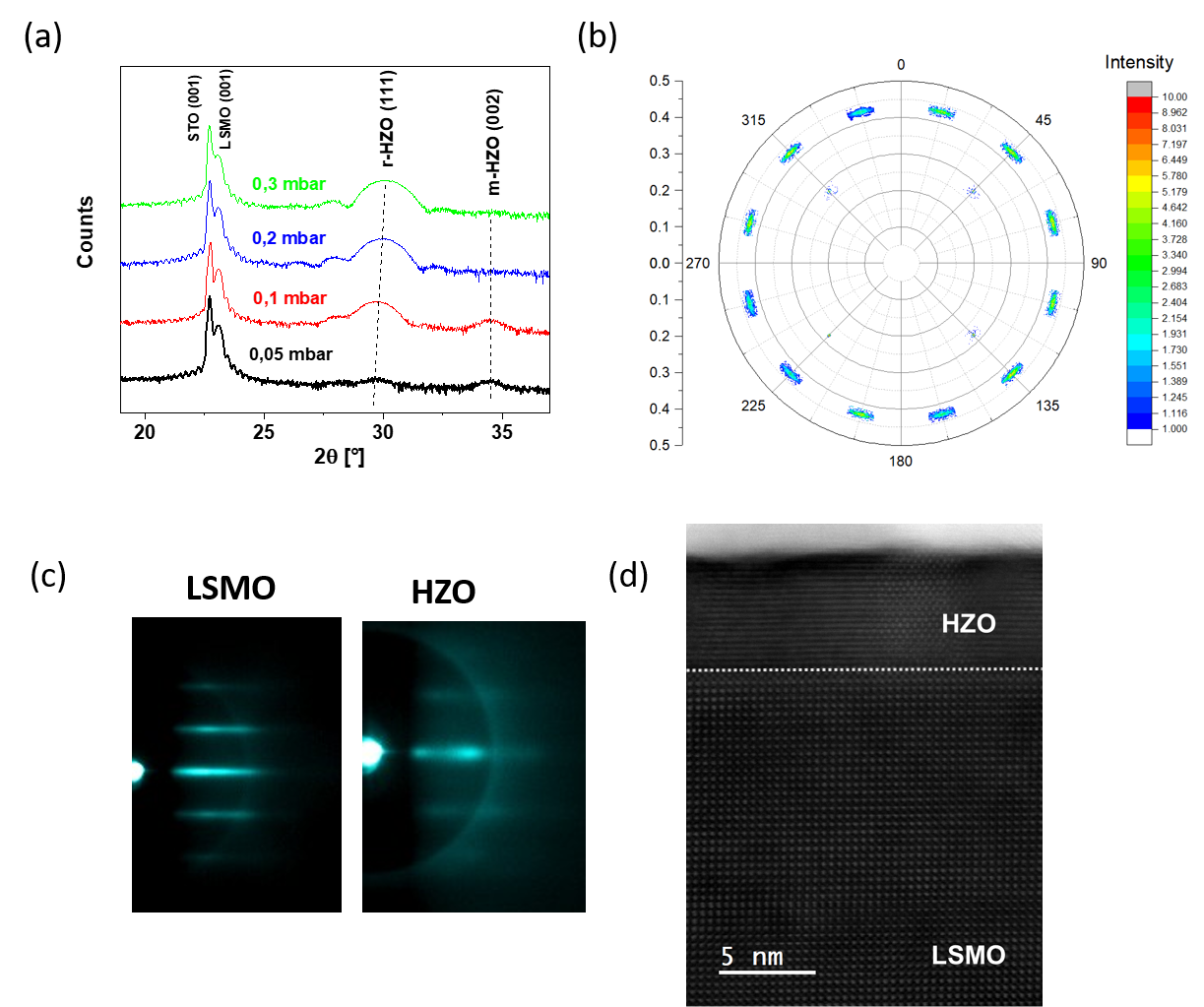}
\caption{(a) $\theta$--2$\theta$ x-ray diffraction patterns of HZO films grown under different oxygen pressures, showing the evolution from mixed monoclinic/rhombohedral phases to single-phase rhombohedral HZO. Laue oscillations around the r-HZO reflection indicate high interface quality and thickness uniformity.
(b) In-plane reciprocal-space map around the HZO (202) reflection as a function of azimuthal angle $\phi$, measured in grazing-incidence geometry. The twelve diffraction maxima at $Q_{\mathrm{in}} \approx 0.43$ r.l.u. are consistent with the multiplicity of the rhombohedral phase. Weak substrate-related in-plane reflections are also visible.
(c) Representative RHEED patterns of the LSMO bottom electrode and HZO layer, indicating smooth epitaxial growth.
(d) Cross-sectional HAADF-STEM image of the HZO/LSMO heterostructure, showing a sharp interface and well-defined crystalline planes.}
\label{fig1}
\end{figure*}

\begin{figure*}[]
\centering
\includegraphics[scale=0.6]{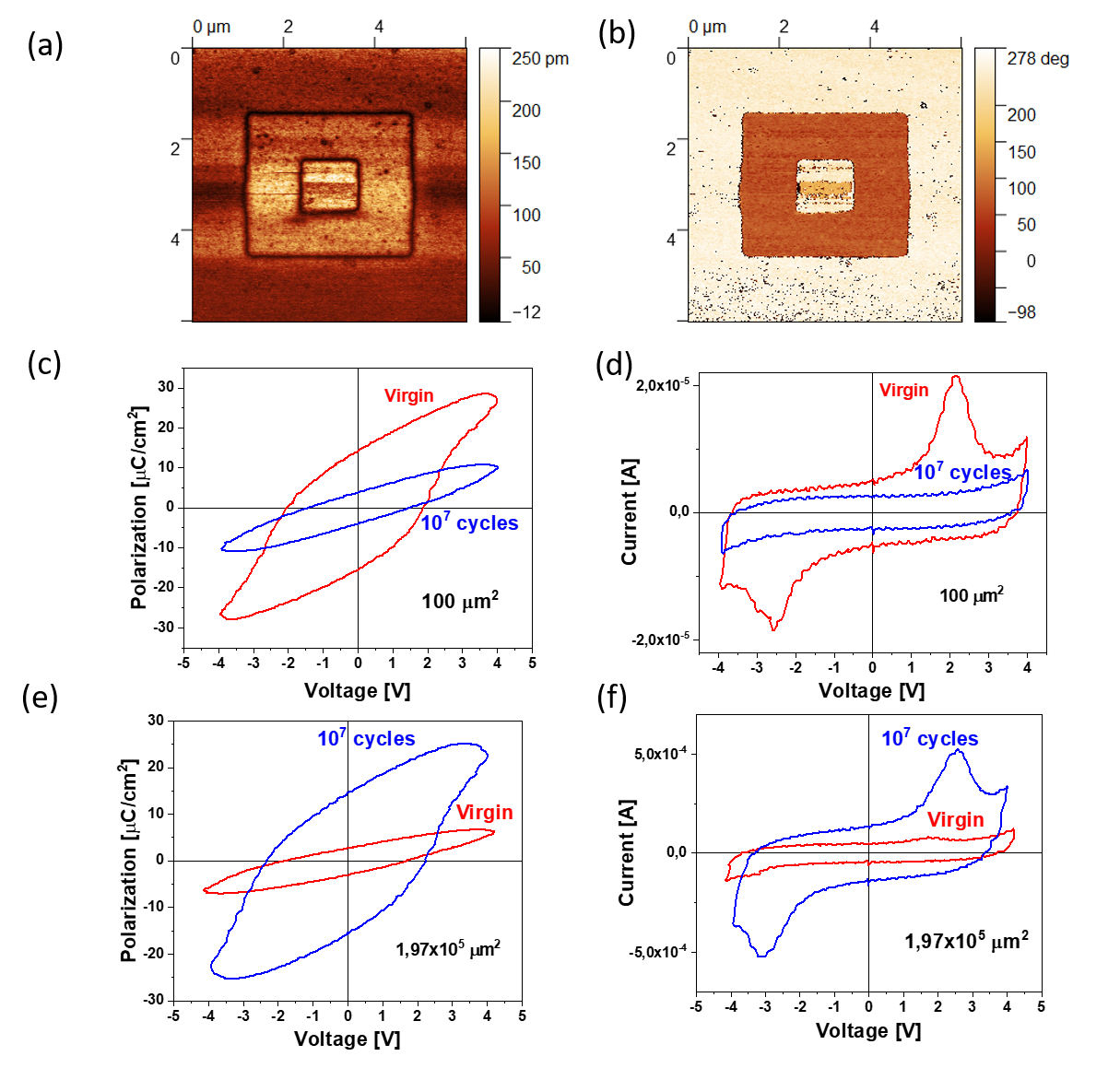}
\caption{PFM amplitude (a) and phase (b) images obtained after writing a square domain pattern on the HZO surface with a biased conductive tip. Polarization--voltage (c) and switching-current (d) measurements for a small-area device, showing wake-up-free ferroelectric switching in the virgin state. (e,f) Corresponding measurements for a large-area device, where a nearly linear virgin response evolves into a pronounced hysteresis with switching currents after repeated cycling, characteristic of wake-up. The slight coercive-voltage difference is consistent with an expected $\approx$10--12\% thickness gradient between distant devices due to stronger plume collimation at high oxygen pressure.}
\label{fig2}
\end{figure*}

The ferroelectric response of the HZO-based devices was further probed electrically through polarization--voltage and switching-current measurements, recorded at room temperature with an aixACCT ferrotester. Figures~2(c) and~2(d) show representative polarization--voltage and current--voltage curves obtained from a 100 $\mu m^2$ small-area device. The polarization loop measured in the virgin state already exhibits a well-defined hysteresis, together with well developed switching currents, indicating wake-up-free ferroelectric switching.

Positive-Up--Negative-Down (PUND) measurements, shown in Fig.~S4, yield a polarization of $\approx 10\,\mu$C/cm$^2$, somewhat lower than values reported for similar epitaxial stacks ($\approx 20\,\mu$C/cm$^2$ \cite{lyu2025IntrinsicFerroelectricHfZrO2}), which may reflect differences in thickness, strain state, or defect distribution. Fatigue measurements, shown in Fig.~S5, reveal that the polarization gradually degrades upon repeated electrical cycling and vanishes after approximately $10^6$ cycles for small-area devices. Fig.~2(c), recorded after the application of $10^{7}$ electrical cycles, confirms the absence of ferroelectric switching after the fatigue test.

In contrast, larger devices display a markedly different evolution. As illustrated in Figs.~2(e) and~2(f), the polarization response of the virgin device with an area of $\approx$ 2x10$^5$ $\mu m^2$ is nearly linear, with no evident switching-current peaks. After repeated electrical cycling ($2.5\times10^{5}$ cycles, see Fig.~S5), however, a clear hysteresis loop emerges and switching-current peaks develop (see Figs.~2(e) and~2(f), recorded after the application of $10^{7}$ cycles). This behavior is characteristic of the wake-up effect commonly observed in hafnia-based ferroelectrics and is typically attributed to the redistribution of oxygen vacancies and the progressive depinning and activation of ferroelectric domains. The distinct evolution of the ferroelectric response in small and large devices indicates that defect landscape and their redistribution during electrical cycling depend on the lateral size of the junctions. 
This size-dependent activation anticipates the emergence of distinct transport regimes.

\begin{figure*}[]
\centering
\includegraphics[scale=0.6]{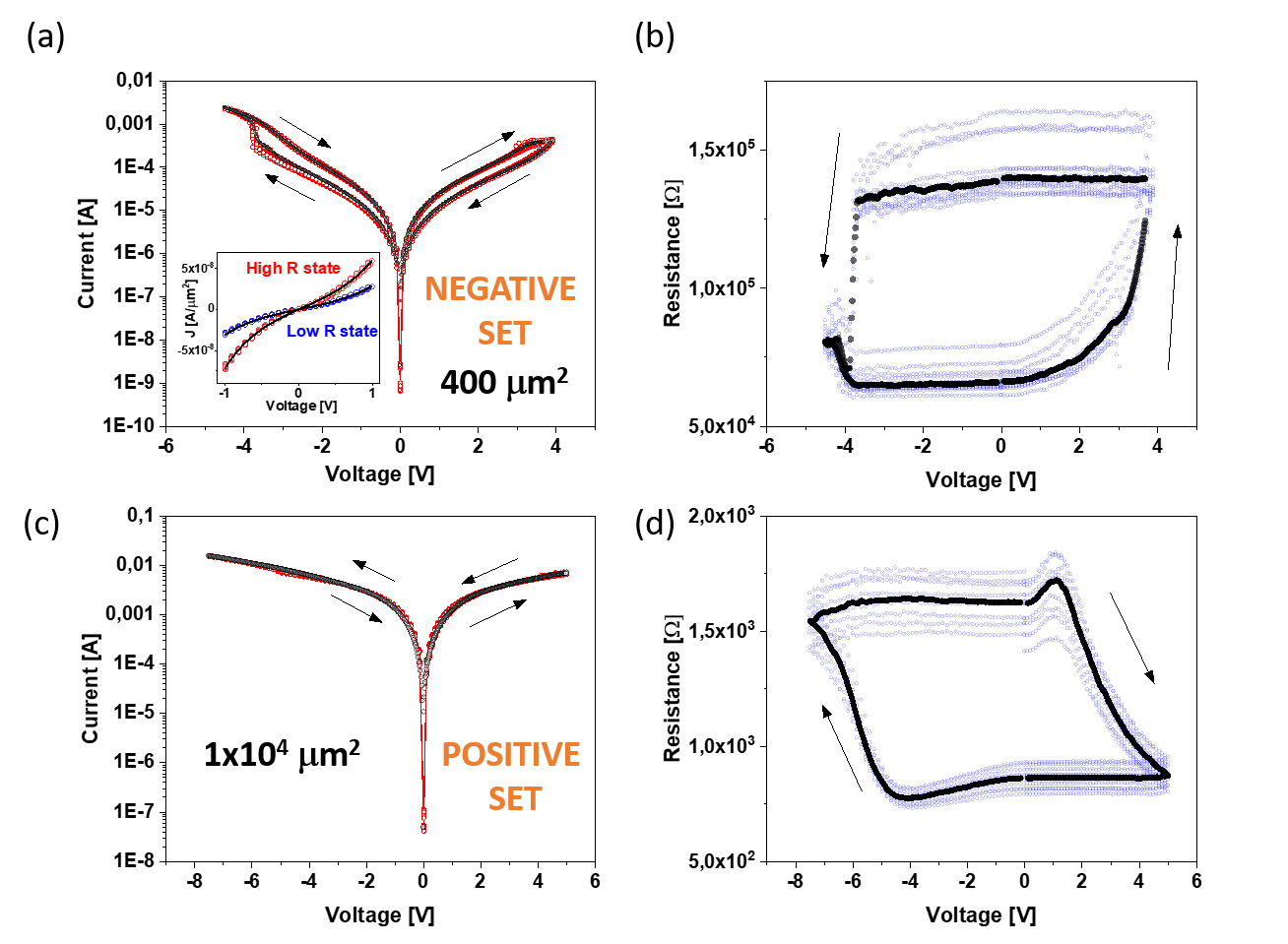}
\caption{(a) Dynamic current--voltage ($I$--$V$) characteristics of a representative small-area device, grown at 0.2 mbar of oxygen, measured over 12 consecutive voltage sweeps. Red symbols show individual cycles, and the black curve corresponds to the median response. The inset shows low-voltage Brinkman fits for both $R_\mathrm{low}$ and $R_\mathrm{high}$ states. (b) Corresponding hysteresis switching loops plotted as resistance versus voltage.
(c,d) Equivalent measurements for a larger-area device.}
\label{fig3}
\end{figure*}

Memristive switching was characterized using a Keithley 2636B source-measure unit (SMU). Sequences of write voltage pulses with progressively increasing amplitude were applied while monitoring the resulting current, yielding dynamic current–voltage curves. In addition, after each write pulse, the device resistance was measured using a 100 mV read pulse, enabling the construction of remanent resistance (HSL) loops. SET and RESET voltages were defined as the write voltages at which the remanent resistance reaches the midpoint between the high- and low-resistance states, $R=(R_{\mathrm{HIGH}}+R_{\mathrm{LOW}})/2$. Representative SET and RESET voltages, together with their statistical variability, are summarized in Table S1 of the Supplementary Material. No conventional electroforming step was required prior to the memristive measurements. Instead, the devices exhibited only a moderate initial resistance stabilization during the first electrical cycle, consistent with a limited electrically induced defect redistribution \cite{Ferreyra25}. Figures~3(a) and 3(b) display the dynamic $I$--$V$ loops and the corresponding remanent hysteresis switching loops (HSLs) \cite{rub_2013,fer_2020, acevedo_2018} for a representative small-area device (400 $\mu m^2$), grown at 0.2 mbar of oxygen. The junction exhibits bipolar switching with a well-defined hysteresis and a SET transition -defined as the resistance change from high ($R_{high}$) to low ($R_{low}$) states- occurring at negative bias applied to the Pt electrode. The remanent HSL confirms the existence of two stable resistance states that can be reversibly accessed by electrical pulses. To further examine the transport mechanism in this regime, the $I$--$V$ curves were analyzed within a Brinkman–WKB description of tunneling through an asymmetric trapezoidal barrier \cite{sul_2019,wei_2019}. In this framework, the current density is given by:

$$
J \cong C\frac{\exp \Bigg\{ \alpha (V) [(\phi_2 - \frac{eV}{2})^{\frac{3}{2}} - (\phi_1 + \frac{eV}{2})^{\frac{3}{2}}] \Bigg\}}{\alpha^2(V)[(\phi_2-\frac{eV}{2})^\frac{1}{2} - (\phi_1 + \frac{eV}{2})^\frac{1}{2}]^2}\times \sinh \Bigg\{ \frac{3}{2} \alpha (V) [(\phi_2 - \frac{eV}{2})^{\frac{1}{2}} - (\phi_1 + \frac{eV}{2})^{\frac{1}{2}}] \frac{eV}{2} \Bigg\}
$$

with

$$
C=-\frac{4em^*m_e}{9 \pi^2 \hbar^3}
$$ and
$$
\alpha (V) \equiv \frac{4d(2m^*m_e)^\frac{1}{2}}{3\hbar(\phi_1 + eV -\phi_2)}
$$


where \(\phi_1\) and \(\phi_2\) are the barrier heights at the two interfaces, \(d\) is the effective barrier thickness and $m^*$ is the electron effective mass, assumed as 0.1$m_e$ \cite{Monaghan2009}. Representative fits for the high- and low-resistance states are shown in the inset of Fig.~3(a) while the extracted parameters from the fittings are displayed in Table~S2. 

The current-voltage characteristics of the negative-SET devices are well described by the Brinkman model for both $R_{high}$ and $R_{low}$, yielding effective barrier thicknesses of $d \approx 3.5$~nm and interface barrier heights in the range of $\sim2$--$4$~eV. The extracted thickness is smaller than the nominal HZO thickness, as expected for an effective tunneling length in an oxide barrier with non-ideal interfaces \cite{sul_2019,wei_2019}.

The transport mechanism was further evaluated against alternative scenarios. A purely thermionic interpretation is unlikely, since the Brinkman analysis shows that the $R_{high}$ to $R_{low}$ transition is dominated by a strong increase in barrier asymmetry rather than by a substantial reduction of the average barrier height, indicating a reshaping of the effective potential profile instead of classical over-the-barrier emission. Likewise, the Fowler--Nordheim representation does not display a clear linear regime within the explored voltage range (Fig.~S6), ruling out Fowler--Nordheim tunneling as the dominant mechanism. Furthermore, temperature-dependent transport measurements together with Poole--Frenkel analysis (Figs.~S12 and S13) indicate that the Brinkman model should be regarded as an effective description of barrier-mediated transport in the distributed conduction regime rather than as evidence of purely elastic direct tunneling. Within this framework, charge transport occurs across an extended asymmetric barrier whose profile is modulated by polarization electrostatics, interfacial dipoles, and electrically active defects, which reshape the barrier and introduce thermally assisted transport channels without producing conventional Poole--Frenkel emission. A possible contribution from reversible interfacial redox (voltammetric) processes recently reported for epitaxial HZO/LSMO heterostructures \cite{nukala_2021} may further contribute to the modulation of the effective barrier, although it cannot by itself account for the positive-SET behavior discussed below. The reduced effective barrier width extracted from the Brinkman analysis is therefore more likely to reflect electronic inhomogeneity than actual thickness variations, consistent with the well-defined x-ray reflectivity oscillations shown in Fig.~S1. A plausible origin is the presence of minority monoclinic or mixed-phase nanoregions \cite{nukala_2021}, which may create lower-barrier hot spots that dominate transport because of the exponential dependence of tunneling probability on barrier height and width.

Larger devices frequently display switching with the opposite polarity, as illustrated in Figs.~3(c) and~3(d) for a representative $1\times10^{4}$~$\mu$m$^{2}$ device. While the dynamic $I$--$V$ curve shows little apparent hysteresis, the HSL clearly reveals bipolar switching with positive SET. These observations indicate the presence of two distinct memristive switching families in our devices, distinguished by their switching chirality. 

We analyzed the area dependence of the switching chirality using a set of 25 devices with different areas and stable resistive switching. Three devices displaying anomalously large $R_{low}$ states, clearly inconsistent with the representative transport ensembles (see Fig.~S7), were excluded from the statistical analysis. Devices were grouped into area bins and we computed the probability of observing positive SET switching as a function of device area, as shown in Fig.~4(a)  

We fitted the probability of positive-SET occurrence with the Poisson nucleation model \cite{ross_2014}
\begin{equation}
P_{\mathrm{SET}+}(A)=P_{\max}\left(1-e^{-\rho A}\right),
\end{equation}
where $\rho$ is the areal density of effective conductive-channel nucleation sites and $P_{\max}$ is the asymptotic probability that at least one such site develops into a stable positive-SET path under the applied electrical protocol. A maximum-likelihood fit yields
$\rho \approx 9.7\times10^{-4}~\mu\mathrm{m}^{-2}$ and
$P_{\max}\approx 0.39$, corresponding to a representative characteristic area
$A^*=\rho^{-1}\approx 1.0\times10^3~\mu\mathrm{m}^2$.

We assessed robustness by parametric bootstrap preserving the area-bin structure (N = 20,000 realizations).
The resulting parameter distributions yield 95\% confidence intervals of 
$\rho \in [4.6 \times 10^{-5}, 5.4 \times 10^{-3}]~\mu\mathrm{m}^{-2}$, 
$P_{\max} \in [0.16, 0.99]$, and 
$A^* \in [1.9 \times 10^2, 2.1 \times 10^4]~\mu\mathrm{m}^2$. The extracted parameters should be interpreted as estimates of the characteristic crossover scale rather than as precise values of a universal nucleation density. Because the available devices cluster around a limited number of lithographic areas, the confidence intervals remain relatively broad. Nevertheless, bootstrap resampling together with sensitivity analyses regarding device selection and area grouping confirm that the characteristic crossover consistently remains of the order of $10^3$ $\mu$$m^2$ (see Supplementary Material).

The broad intervals reflect the limited size of the present ensemble, so the model should be interpreted as a minimal statistical description of the observed monotonic trend rather than as a precise determination of microscopic parameters.

Overall, the increasing probability of positive-SET events with junction area is consistent with the statistical emergence of localized conductive-channel nucleation sites.

\begin{figure*}[]
\centering
\includegraphics[scale=0.55]{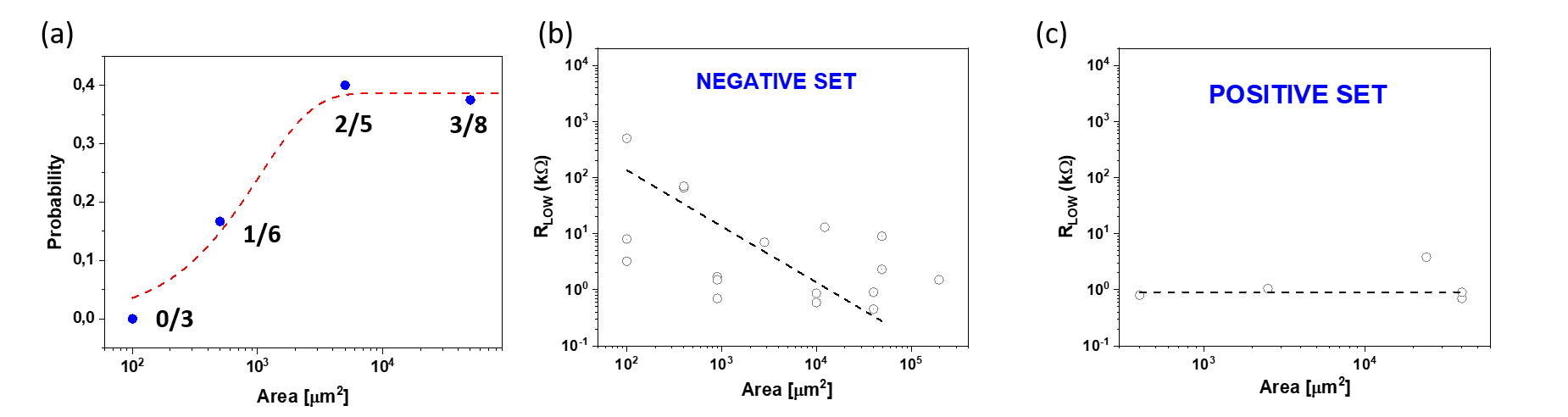}
\caption{(a) Probability of positive SET switching as a function of device area, for devices grown at 0.2 mbar of oxygen. Symbols correspond to experimental data and the dashed line to the maximum-likelihood Poisson fit. Labels indicate positive-SET events over the total number of devices in each bin. (b) Area dependence of the low-resistance state $R_{\mathrm{low}}$ for negative-SET devices. Open symbols show individual devices and the dotted line the fit $R_{\mathrm{low}} = C/A$.
(c) Area dependence of $R_{\mathrm{low}}$ for positive-SET devices. Open symbols show individual devices and the dotted line the median trend. Negative-SET devices approximately follow $1/A$, whereas positive-SET devices remain area-independent.}
\label{fig4}
\end{figure*}

To evaluate the reproducibility of the observed transport regimes under different growth conditions, we additionally investigated devices fabricated with a slightly higher oxygen deposition pressure (0.3 mbar instead of 0.2 mbar, while keeping the remaining growth parameters unchanged). Under these conditions, nearly all measured devices exhibited positive-SET behavior, together with higher resistance levels and an enlarged ON/OFF resistance window, while preserving robust ferroelectric switching (see Supplementary Material, Figure S11). These observations further support the interpretation that the statistical distribution of transport regimes is governed by the defect landscape established during film growth, and suggest that defect engineering provides a viable route to optimize the memristive response of epitaxial HZO devices.

To further elucidate the transport mechanisms governing the ON state, we analyzed the dependence of $R_{\mathrm{low}}$ on device area for both switching polarities. Figs.~4(b) and (c) summarize the results for the negative and positive SET branches, respectively. For the negative SET polarity [Fig.~4(b)], $R_{\mathrm{low}}$ exhibits a dependence on device area with significant device-to-device variability, consistent with electrical transport with a significative stochastic component. For the smallest areas, $R_{\mathrm{low}}$ reaches values up to $\approx$ $50\,\mathrm{k}\Omega$; as the device area increases, the resistance decreases systematically and approaches  values of the order of $1$--$1.5\,\mathrm{k}\Omega$ for the largest devices. The large dispersion observed in different devices with similar areas is consistent with a spatially distributed transport mechanism across the ferroelectric barrier, where local inhomogeneities modulate the effective tunneling conductance while preserving an overall scaling close to 1/A.
In contrast, the positive SET polarity [Fig.~4(c)] shows different behavior. In this case, the low-resistance state remains nearly independent of device area across the entire range investigated, with most devices converging to a characteristic value of $R_{\mathrm{low}}\sim1\,\mathrm{k}\Omega$. 

The absence of area scaling suggests that the conduction is dominated by localized conductive paths~\cite{saw_2008}, likely associated with defect-mediated transport or oxygen-vacancy aggregation. Once such a localized channel is formed, the device resistance is determined by the properties of this path rather than by the total junction area. These localized conductive paths are interpreted as preferential transport paths generated by defect reorganization, rather than as permanent soft-breakdown filaments.

A simple estimate based on the write-pulse protocol indicates that the distributed transport regime requires switching energies about three orders of magnitude lower than the localized-channel regime, highlighting its potential for ultra-low-power neuromorphic hardware. The observed ON/OFF ratios ($\sim$2 for both chiralities) are comparable to those commonly reported for epitaxial HZO ferroelectric memristors without electroforming \cite{sul_2019, Ferreyra25} and can be substantially increased by a slight adjustment of the oxygen pressure during growth (see Figure S11 of the Supplementary Material) while preserving robust ferroelectric switching. The relatively limited ON/OFF ratio observed in the negative-SET regime, compared with optimized perovskite FTJs \cite{Wen2020}, supports the coexistence of polarization-modulated barrier transport with additional defect-assisted conduction channels rather than an ideal FTJ regime. In the positive-SET regime, the absence of area scaling together with similarly low ON/OFF ratios suggests transport through localized conductive paths undergoing only partial disruption upon RESET, consistent with reversible oxygen-vacancy rearrangement rather than permanent filamentary breakdown \cite{saw_2008}.

Devices with areas well below the characteristic area $A^*$ are unlikely to contain a nucleation site and therefore predominantly remain in the distributed barrier-mediated transport regime, whereas larger junctions increasingly sample the nucleation-site distribution and develop localized conductive channels. This model captures the statistical crossover shown in Fig.~4(a) and directly links device area to the dominant transport mechanism. The same physical picture is also consistent with the size-dependent wake-up behavior: large-area devices develop switching only after electrical cycling, whereas small-area devices are already active in the virgin state. Since the memristive characterization inherently involves repeated bipolar electrical cycling, large-area devices naturally evolve toward a wake-up-equivalent state during the memristive measurement protocol, as previously demonstrated in Ref.~\cite{Ferreyra25}. Although direct microscopic evidence of oxygen-vacancy reorganization was not obtained in the present work, this mechanism provides the most plausible explanation for the observed transport evolution, consistent with the correlation between transport crossover and wake-up \cite{Martemucci2025}, the sensitivity to growth conditions, and the temperature-dependent transport measurements \cite{Schulman2024}.


In summary, lateral device area statistically selects the dominant transport mechanism in epitaxial HZO devices, driving a crossover from barrier-controlled tunneling to localized conduction. This transition is quantitatively described by a Poisson nucleation model and correlates with the onset of ferroelectric wake-up, supporting a common oxygen-vacancy redistribution origin. These results provide a design guideline linking device geometry and memristive response in hafnia ferroelectric devices.

\textbf{Supplementary Material}

See the supplementary material for additional structural and electric characterization and for the identification of outlier devices.

\textbf{Acknowledgements}

Some of the experiments reported here were carried out within the NFFA-Europe access proposals ID478 and ID579 (2023), funded by the European Union’s Horizon 2020 research and innovation programme under Grant Agreement No. 101007417, and the CNRS-IRP action PhenomeNaS. We also acknowledge support from EU-H2020-RISE project MELON (Grant No. 872631). We also thank financial support from the Instituto de Tecnología (INTEC) of the Universidad Argentina de la Empresa (UADE), project code A26T63. The financial support from the Groningen Cognitive Systems and Materials Center (CogniGron) and the Ubbo Emmius Foundation of the University of Groningen is appreciated. We thank M. Sarott, from the University of Groningen, for his assistance with the PFM measurements.


\textbf{AUTHOR DECLARATIONS}

\textbf{Conflict of Interest}

The authors have no conflicts to disclose.

\textbf{DATA AVAILABILITY}

The data that support the findings of this study are available
from the corresponding author upon reasonable request.

\bibliography{bibliography.bib}

\end{document}